\documentclass[prl,twocolumn,showpacs]{revtex4}

\usepackage{graphicx}
\usepackage{amsmath}
\usepackage{epstopdf}
\usepackage{amsmath}
\usepackage{amsfonts}

\usepackage{epsfig} 

\usepackage{mathtools}% http://ctan.org/pkg/mathtools

\usepackage{color}

\begin{document}

\title{Exact  overlaps in the Kondo problem}

\author{Sergei L.\ Lukyanov${}^1$, Hubert Saleur$^{1,2,3}$, Jesper L.\ Jacobsen$^{4,5}$ and Romain Vasseur$^{6,7}$}

\affiliation{${}^1$Physics Department, Rutgers University, Piscataway, NJ, USA}
\affiliation{${}^2$Department of Physics and Astronomy,
University of Southern California, Los Angeles, CA 90089, USA}
\affiliation{${}^3$Institut de Physique Th\'eorique, CEA Saclay,
91191 Gif Sur Yvette, France}
\affiliation{${}^4$LPTENS, \'Ecole Normale Sup\'erieure, 24 rue Lhomond, 75231 Paris, France}
\affiliation{${}^5$Universit\'e Pierre et Marie Curie, 4 place Jussieu, 75252 Paris, France}
\affiliation{${}^6$Department of Physics, University of California, Berkeley, CA 94720, USA}
\affiliation{${}^7$Materials Science Division, Lawrence Berkeley National Laboratories, Berkeley, CA 94720, USA}

\date{\today}

\begin{abstract}
It is well known that the ground states of  a Fermi liquid with and without a single  Kondo impurity have an overlap which decays as a power law of the  system size, expressing the Anderson orthogonality catastrophe. Ground states with two different values of the Kondo couplings have, however, a finite overlap in the thermodynamic limit. This overlap, which plays an important role in quantum quenches for impurity systems, is a universal function of the ratio of the corresponding Kondo temperatures, which is not accessible using perturbation theory nor the Bethe ansatz. Using a strategy 
based on the integrable structure of the corresponding quantum field theory,
we propose an exact formula for this overlap, which we check against extensive density matrix renormalization group calculations. 
\end{abstract}


\maketitle

\paragraph{Introduction.}

The Anderson orthogonality catastrophe (AOC) is one of the cornerstones of modern many body physics. In its simplest formulation, this ``catastrophe'' states that the ground states of two Fermi seas with different local scattering potentials become (if the orthogonality exponent is non zero),  orthogonal in the thermodynamic limit. This fact has many important consequences, and is at the root of the physics of Mahan excitons~\cite{Mahan}, the Fermi edge singularity in absorption spectra~\cite{Nozieres}, the non linear $I_V$ characteristics in quantum dots, or the Kondo effect~\cite{Hewson} in magnetic alloys. More recently, AOC has played a central role in understanding the post quench dynamics induced by optical absorption in quantum dots tunnel-coupled to Fermi seas~\cite{Exp1,Exp2}. 

The simplest manifestation of AOC occurs in the case of a free Fermi  
sea involving a single channel of non interacting electrons that experience 
two different local scattering potentials. If the corresponding phase 
shifts {\sl at the Fermi energy} are 
$\delta_F^{(1)},\delta^{(2)}_F$, a 
simple argument~\cite{Anderson} shows that the scalar product of the ground states vanishes as
\begin{equation}
\left\langle\, \Omega_{2}\,|\,\Omega_{1}\,\right\rangle\propto N^{-(\delta^{(1)}_F-\delta^{(2)}_F)^2/2\pi^2}\label{eq1},
\end{equation}
where $N$ is the total number of electrons. AOC occurs as well in interacting systems. 
In the $k$-channel Kondo problem for instance, it is known that the scalar product of  the system 
with and without a Kondo impurity behaves as 
$\langle\, \Omega(J)\,|\,\Omega(J=0)\,\rangle\propto 
N^{-d_K} $ where \cite{AffleckLudwig} $d_K=\frac{3}{ 4(k+2)}$
and $J$ is the  (antiferromagnetic) Kondo coupling. The simplest one-channel case, 
to which we will restrict in this Letter,  corresponds then to $d_K={1\over 4}$. 
In this case, the orthogonality of the  ground states  expresses  the fact  that at very low energy, 
spin up and spin down electrons see a phase shift  
of $0$ (resp.\ ${\pi\over 2}$) with  zero (resp. non zero) 
Kondo coupling.  An easy generalization of this argument gives 
the exponent in the anisotropic Kondo case as well: 
in the Toulouse limit in particular,  $d^{(\rm Tou)}_K={1\over 8}$. 
No such simple Fermi liquid calculation  exists for $k>1$, 
and  sophisticated techniques have to be used to calculate the overlap, 
such as integrability or conformal invariance. In the latter set-up, the 
orthogonality exponent  
$d_K$ is interpreted as the scaling dimension  of a boundary 
condition changing operator~\cite{AffleckLudwig}.  
Note that such exponents are directly related to the power law tail in  
the so-called work distribution~\cite{WorkSilva,HKXray} for quenches when a coupling is 
suddenly turned on, such  as those studied in Refs.~\cite{Exp1,Exp2} in the Kondo case.

The ground state overlap exemplifies the {\sl non perturbative} quantities
occurring in quantum impurity problems. 
An interesting variant is provided by the overlap 
$\langle\, \Omega_2\,|\,\Omega_1\,\rangle$ 
between ground states corresponding to two different non-vanishing Kondo couplings $J^{(1)},J^{(2)}$.  
This overlap is not expected to vanish when both $J^{(1)},J^{(2)}\neq 0$, 
even in the thermodynamic limit. This is because, for any non zero Kondo coupling, 
fermions at very low energy now see the same phase shift of ${\pi\over 2}$. 
Nevertheless, this overlap is non trivial, even in the non interacting Toulouse limit, 
because it is  determined by the behavior of 
the whole Fermi sea, and not just what happens at the Fermi energy. 
This overlap is also {\sl non perturbative}: any attempt to calculate it by expanding in 
$J^{(1)},J^{(2)}$ is plagued by infrared divergences precisely  
because of the AOC.  
Overlaps such as  $\langle\, \Omega_2\,|\,\Omega_1\,\rangle$ 
arise  in quantum quenches  where one suddenly changes the Kondo coupling 
$J^{(1)}\mapsto J^{(2)}$. 
The system then has a finite probability of remaining in the ground state at large times, 
which translates into a delta function in the corresponding work distribution \cite{WorkSilva}: 
this probability is precisely the square modulus $P_{1\mapsto 2}=
|\langle\, \Omega_2\,|\,\Omega_1\,\rangle|^2$, 
and it could be measured in optical absorption experiments realizing such quantum quenches~\cite{Exp1,Exp2}.

Now, the Kondo problem exhibits universal properties at energy scales much smaller than the bandwidth $D$. 
In this limit, physical quantities depend only on the temperature, 
the magnetic field, and a crossover scale which encodes the Kondo coupling $J$ 
(see their precise relationship below) -- the Kondo temperature $T_K$. 
Different (proportional) definitions of $T_K$ exist, but this will not matter for us. 
Indeed, provided $T^{(1)}_K,T^{(2)}_K \ll D$, scaling arguments show 
that the overlap becomes a universal function  of the ratio
\begin{equation}
\langle\, \Omega_2\,|\,\Omega_1\,\rangle =
F\big(\,T^{(1)}_K/T^{(2)}_K\,\big)=F\big(\,T^{(2)}_K/T^{(1)}_K\,\big)\ .
\label{eqUniversalFunction}
\end{equation}
 In this Letter, we obtain an exact formula for this quantity,  
which we also check with extensive Density Matrix Renormalization Group (DMRG) calculations. 

\paragraph{Anisotropic Kondo model.}

The anisotropic Kondo problem is initially formulated as a three dimensional problem of non interacting 
fermions coupled to a local magnetic impurity. 
After a spherical waves  decomposition, only the $s$-channel interacts with the impurity, 
and the problem can be transformed into 
one dimensional gapless fermions on the half line (the radial coordinate) coupled to a spin at the origin. 
``Unfolding'' the half line one obtains a problem of chiral fermions with   
\begin{eqnarray}
{\boldsymbol H}&=& -{\rm i}\,v_{F}\sum_{\alpha=\uparrow,\downarrow}\int_{-\infty}^\infty {\rm d}x\, \psi_{\alpha}^{\dagger}\partial_{x}\psi_{\alpha}\\
&+&J \left(\,j^+(0)\,\sigma^-+j^-(0)\,\sigma^+\,\right)+J_z j^z(0)\,\sigma^z\,,\nonumber
\end{eqnarray}
where the spin currents are 
$j^+=\psi_\uparrow^\dagger\psi_\downarrow,j^-=\psi^\dagger_\downarrow\psi_\uparrow,j^z=
\psi^\dagger_\uparrow\psi_\uparrow-\psi^\dagger_\downarrow\psi_\downarrow$. 
We bosonize the fermionic fields 
$\psi_\sigma \sim \mathrm{e}^{{\rm i} \sqrt{4 \pi} \phi_\sigma}$~\cite{Giamarchi}, 
which allows us to separate charge and spin modes 
$\phi_{c/s}=(\phi_\uparrow \pm \phi_\downarrow)/\sqrt{2}$. 
The charge boson decouples, and the  interacting part  involves only the spin boson $\phi=\phi_s$. 
After a canonical transformation, ${\boldsymbol H} \to 
{\boldsymbol U}^\dagger {\boldsymbol H} {\boldsymbol U}$ 
with ${\boldsymbol U}=\exp({\rm i}J_z \phi(0)\sigma_z)$, 
one can then rewrite the Hamiltonian as
\begin{equation}
{\boldsymbol H}=\int_{-\infty}^\infty {\rm d}x\, 
(\partial_x\phi)^2+J \left({\rm e}^{{\rm i}\beta\phi(0)}\sigma^-+
{\rm e}^{-{\rm i}\beta\phi(0)}\sigma^+\right)\label{K1},
\end{equation}
with $\beta=\sqrt{8\pi}-2J_z$ and the equal time commutation relations $[\phi(x),\phi(x')]=
{{\rm i} \over 4}\,\hbox{sign }(x-x')$. 
The scaling dimension of the perturbation is ${\beta^2\over 8\pi}=
\left(1-{J_z\over \sqrt{2\pi}}\right)^2\equiv {\xi\over \xi+1}$ 
where the last equality defines the coupling constant $\xi$. 
The Kondo temperature in this framework varies as $T_K\propto J^{\xi+1}$.

   \begin{figure}[t!]
\includegraphics[width=\columnwidth]{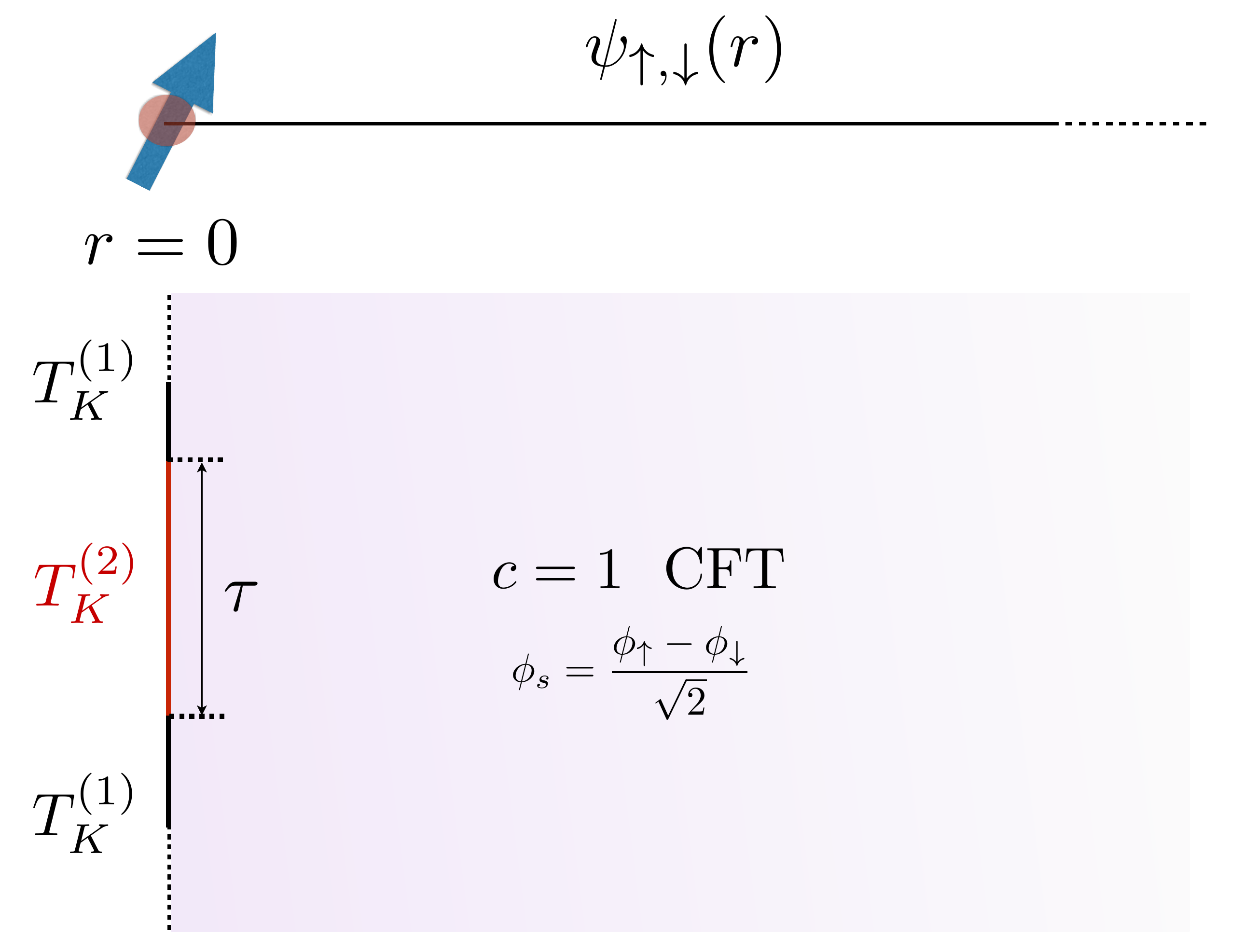}
\caption{The overlap $\left|  \langle\, \Omega_2\,|\,\Omega_1\,\rangle \right|$ 
can be extracted from the partition 
function of the system with the insertion of a 
Kondo impurity and two different values of the coupling. 
In this picture, the boundary condition 
corresponds to the spin impurity while the bulk describes a 
critical statistical mechanics problem associated with the spin mode $\phi_{s}$. }
\label{figure1}
\end{figure}

\paragraph{Perturbative results.}

It is first natural to try to evaluate the universal 
function~\eqref{eqUniversalFunction} using perturbation theory. 
To this end, we fold the chiral problem~\eqref{K1} to obtain a non-chiral boson on 
the half line $( - \infty, 0]$, scattering off the spin impurity at $x=0$. 
We then map this $(1+1)D$ quantum impurity system onto a $2D$ classical statistical 
mechanics problem in the half-plane, critical in the bulk (corresponding to the $c=1$ free boson theory), 
with the impurity now acting as a boundary condition (see Fig.~\ref{figure1}). We then calculate the partition function 
${\cal Z}\big(J^{(1)},J^{(2)}\big)$ of a half-infinite system with 
boundary condition corresponding to the Kondo temperature 
$T^{(1)}_K$ 
everywhere except on a part of the boundary of length $\tau$ where the boundary field is 
taken to correspond to $T^{(2)}_K$.  
It gives a term linear in imaginary time (corresponding to a boundary free energy contribution), 
a term exponential in imaginary time (corresponding to excited states propagating along the boundary), 
and a term of order one which can be seen to be 
$ |\langle\, \Omega_2\,|\,\Omega_1\,\rangle|^2$ in the Hamiltonian formalism. 
 
Expanding the overlap $\left|  \langle\, \Omega_2\,|\,\Omega_1\,\rangle \right|^2$ 
 in $J^{(1)}-J^{(2)}$ 
from the ratio ${\cal Z}(J^{(1)},J^{(2)})/{\cal Z}(J^{(1)},J^{(1)})$ is
extremely complicated, 
since the two-point function of the boundary perturbation in~\eqref{K1} is not known in general.
At the Toulouse point ($\xi = 1$), however, the perturbation can be refermionized,
so the spin-spin propagator at finite value of $J$ is easily found, 
and expanding the partition function yields~\cite{SupMat}
 \begin{equation}
|\langle\, \Omega_2\,|\,\Omega_1\,\rangle|_{\xi=1}=1-
\frac{\alpha_{12}^2}{ 8\pi^2} + {\cal O}(\alpha_{12}^4)\, , \label{ratioresi}
 \end{equation}
where  
$\mathrm{e}^{\alpha_{12}}=T^{(2)}_K/T_K^{(1)}$. 
Even for this non-interacting case, going beyond this first order expansion becomes quickly involved, 
and capturing the full behavior of the function~\eqref{eqUniversalFunction} seems hopeless. 
  
\paragraph{Semi-classical analysis.}

The overlap can also be calculated perturbatively in
the semiclassical limit, where $\xi \simeq \frac{\beta^2}{8 \pi} \ll1$. 
In this case, it is convenient to implement yet another canonical transformation, and bring the Hamiltonian into the form
\begin{equation}
{\boldsymbol H}=
{{1\over 2}}\int_{-\infty}^0{\rm d} x \left((\partial_x\Phi)^2+(\partial _t\Phi)^2\right)+J \sigma^x+
{ {\beta\over 4}}\, \partial_t \Phi(0)\sigma^z.\label{semiclass}
\end{equation}
Using perturbation theory in $\beta$, we now calculate the partition 
function ${\cal Z}$ in imaginary time of a system 
with two different values of $J$ as shown in Fig.~\ref{figure1}. 
The leading contribution comes from the configuration 
where $\sigma^x=-1$ everywhere but between a pair of insertions, 
spaced by $\tau$, of the $\partial_t \Phi(0)\sigma^z$ term. 
Discarding terms that depend on $\tau$ and encode the non universal boundary free energy, we find~\cite{SupMat}
 \begin{equation}
| \langle\, \Omega_2\,|\,\Omega_1\,\rangle| =
1+{\xi\over 2}
\left(1-{\alpha_{12}\over 2}\coth{\alpha_{12}\over 2}\right) + {\cal O}(\xi^2) \label{ratiores}\, .
 \end{equation}
 Once again, going beyond this first order is extremely involved, and there is, 
in particular, no chance to capture the crossover between the two extreme behaviors, $T_K^{(1)}\sim T_K^{(2)}$ and 
$T_K^{(1)}\gg T_K^{(2)}$.  
 
\paragraph{Exact results from integrability.} 

For many other questions in the Kondo problem -- such as the study of thermodynamics 
properties~\cite{TsvelickWiegmann,AndreiLowenstein}, 
correlation functions~\cite{Lesage}, quantum quenches~\cite{PRLQuench} 
or entanglement~\cite{Vasseur} -- non perturbative 
techniques have led to analytic 
expressions in the crossover regions, when 
the physical scale of interest (temperature, magnetic field, etc) is comparable with $T_K$. 
Although exact Bethe ansatz wave functions are in principle known for different values of $T_K$,  
overlaps such as $\langle\, \Omega_2\,|\,\Omega_1\,\rangle$, 
have however  proven, so far,  
impossibly hard to calculate directly. 
We report here another approach to the problem based on an axiomatic determination of the overlaps 
directly in the field theory limit. 
This approach is similar in philosophy to the  $S$-matrix bootstrap  from Ref.~\cite{Zam}.
% what has been done for matrix elements of local observables - the so called 
%form-factors program~\cite{Smirnov}. 
We give  the relevant details in the supplementary material, and move directly to the main result. 

 \begin{figure}[t!]
\includegraphics[width=1.0\linewidth]{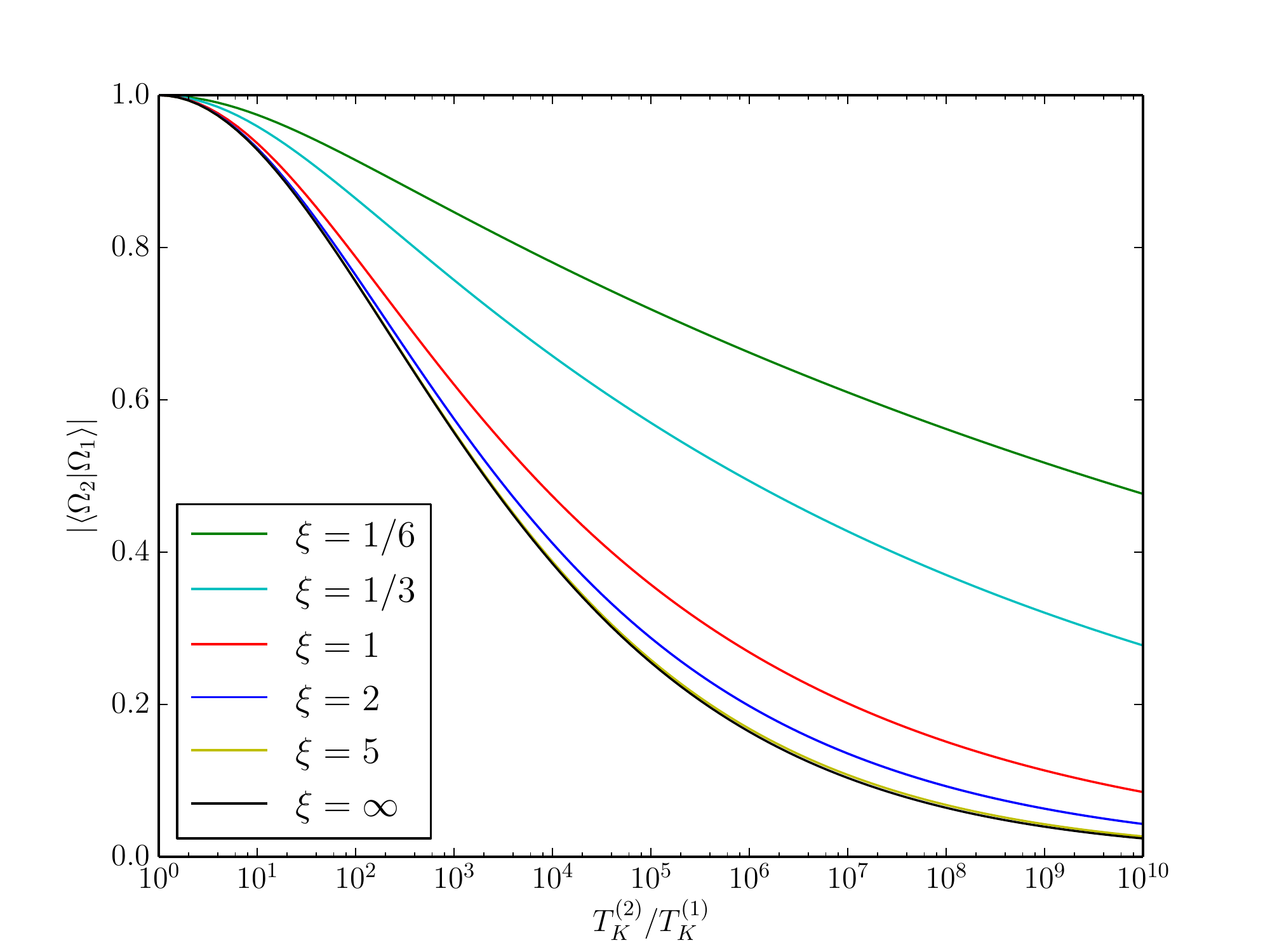}
\caption{Theoretical result for the overlap for various 
anisotropies as a function of the ratio $T_K^{(2)}/T_K^{(1)}$. 
The isotropic Kondo problem then corresponds to $\xi = \infty$. Note the extreme values on the $x$-axis.}
\label{figure4}
\end{figure}
%
%
%The apparition of two different $R$ matrices in the problem is profoundly related with the fact that 
%the quantum symmetries of the continuum Kondo theory and its lattice regularizations such as the 6 vertex model with inhomogeneous spectral parameters \cite{ReshSal} are different.

 \begin{figure*}[t!] 
%\begin{figure}[t]
 \centering
 
 \includegraphics[width=0.49\textwidth]{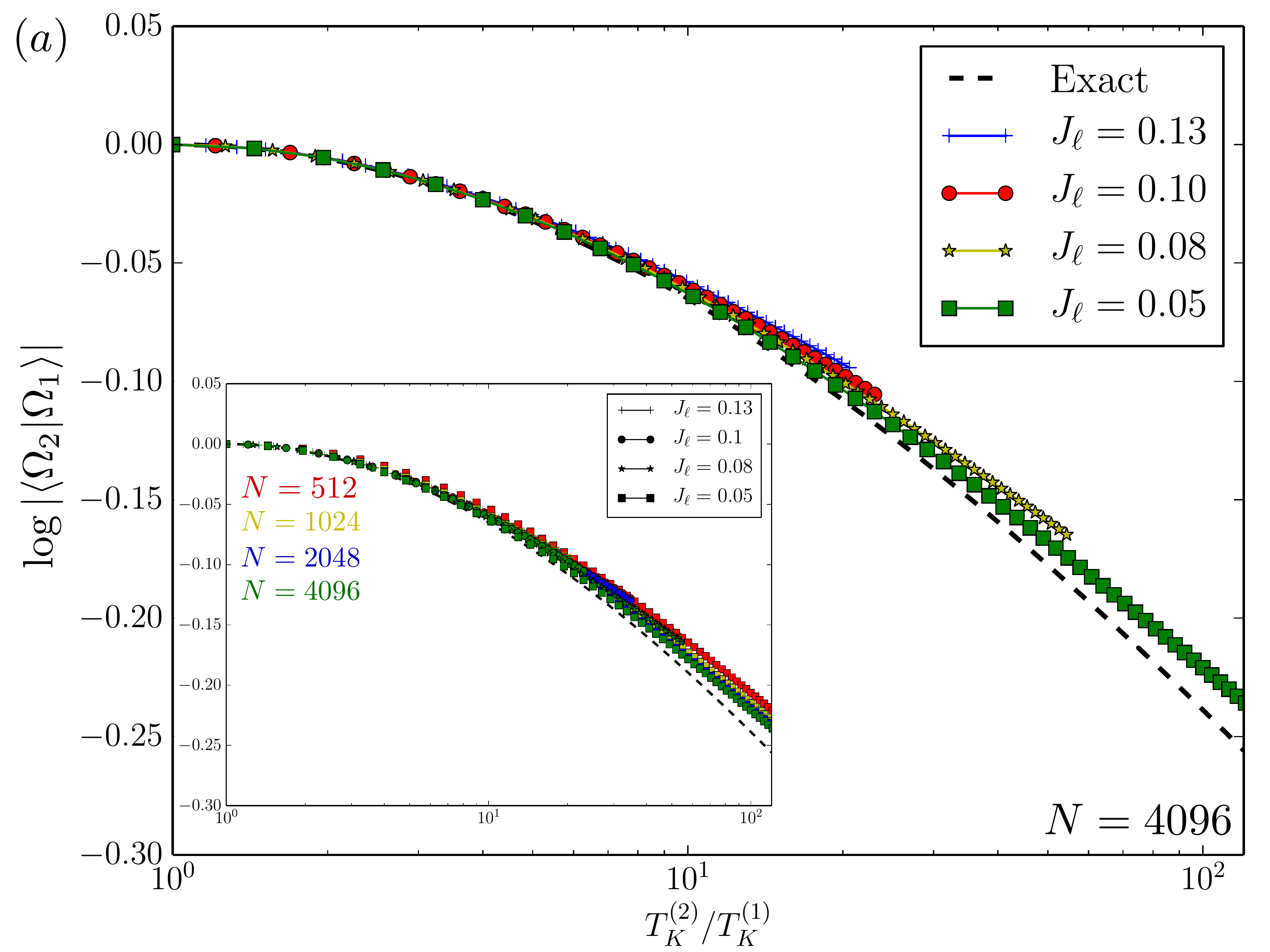}
 \includegraphics[width=0.49\textwidth]{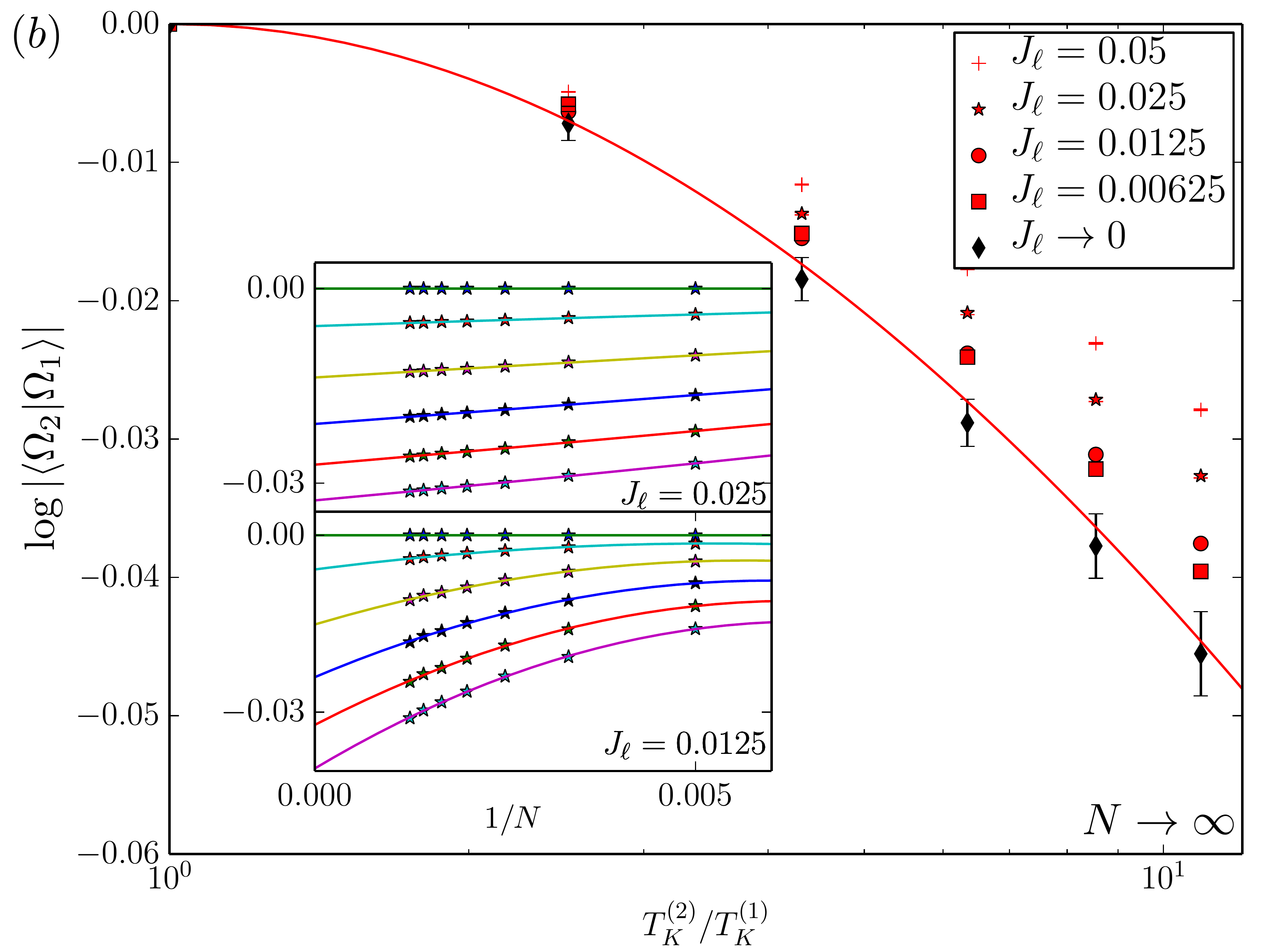}

 \caption{Numerical checks of the main result~\eqref{mainres}. Note that the comparison 
between numerical and field theory 
results does not involve any free parameter. (a) Measures 
of the overlap in the free-fermion Toulouse  
case $\xi = 1$, modeled by a spinless non-interacting resonant level, 
tunnel-coupled with parameter $J_\ell$ to two 
metallic reservoirs with $N=4096$ sites each. 
The dashed line is the analytical result, with 
$T_K\propto J_\ell^{2}$. Inset: Finite size scaling.   (b) DMRG results for the overlap in 
the interacting case $\xi = 1/3$, modeled by an XXZ 
spin chain with anisotropy $\Delta=-1/\sqrt{2}$ 
and $N$ sites, with an extra site at the edge 
characterized by a weak link $J_\ell$, corresponding to the impurity. 
The numerical results for $N=200, \dots, 800$ are extrapolated in the 
thermodynamic limit $N \to \infty$. 
The red line is the analytical result, 
with $T_K\propto J_\ell^{4/3}$, and the black symbols correspond to 
the extrapolation $J_\ell \to 0$. Inset: Examples of finite size 
extrapolations for fixed $J_\ell \equiv J_\ell^{(1)}$ and different values of  $J_\ell^{(2)}$. }
  \label{fig:Numerics}
\end{figure*}
 
%
%The minimal solution to the equations satisfied by the $T$ matrices leads to 
%%
%\begin{equation}
%F(\alpha_{12})\equiv \langle T_\pm(\alpha_2)T_\mp(\alpha_1)\rangle=\pm(1+\xi) ~{\sinh{\alpha_{12}+i\pi\over2(1+\xi)}\over \sinh{\alpha_{12}+i\pi\over 2}}~
%G'(\alpha_{12})
%\end{equation}
%%
%This quantity is normalized such that 
%%
%\begin{equation}
%F(-i\pi)=\langle T_\pm(\alpha)T_\mp(\alpha+i\pi)\rangle=1
%\end{equation}
%%
%Here,
%Define first 
%%
%\begin{equation}
%G'(\alpha)=\exp\left[-\int_0^\infty {dt\over t} {\sinh^2 t(1-{i\alpha\over\pi}) \over \sinh 2t\cosh t}{\sinh t\xi\over \sinh t(\xi+1)}\right]\label{Gfunct}
%\end{equation}
%%
%We can now go back to our initial goal of calculating the overlaps. In the language of quantum Jost operators, all we have to do is take the product $T(J)Q(J')$ where now $T$ and $Q$ have different arguments. The final result is now
%%
%\begin{eqnarray}
%\left\langle\Omega\left(T_K^{(2)}\right)|\Omega\left(T_K^{(1)}
%\right)\right\rangle=\langle T_\pm(\alpha_2+i\pi)T_\mp(\alpha_1)\rangle=\nonumber\\
%(1+\xi) ~{\sinh{\alpha_{12}\over2(1+\xi)}\over \sinh{\alpha_{12}\over 2}}~
%G(\alpha_{12}),~~~e^{\alpha_{12}}={T_K^{(1)}/T_K^{(2)}}\label{mainres}
%\end{eqnarray}
%%
%with $G(\alpha)=G'(\alpha-i\pi)$. 

We find that the  overlap is given by
\begin{equation}
%\label{final}
 \langle\, \Omega_2\,|\,\Omega_1\,\rangle =
(\xi+1) ~{\sinh{\alpha_{12}\over2(\xi+1)}\over \sinh{\alpha_{12}\over 2}}~
g_\xi(\alpha_{12}) 
\label{mainres}
\end{equation}
with 
\begin{equation}
\label{gxi}
g_\xi(\alpha)=\exp\left(\int_0^\infty {{\rm d}t\over t} {\sin^2 (\alpha t/\pi) \over \sinh 2t\cosh t}{\sinh t\xi\over \sinh t(\xi+1)}\right)\,,
\end{equation}
where we recall that $\mathrm{e}^{\alpha_{12}}=T^{(2)}_K/T_K^{(1)}$. 
See Fig.\,\ref{figure4} for a plot of this exact solution, illustrating the 
variation of the overlap with the anisotropy, as well as the incredibly 
large values of the ratio $T_K^{(2)}/T_K^{(1)}$ necessary to 
bring this overlap down to the $10^{-1}$ or less. 
It  is worth mentioning here that
the function $g_\xi(\alpha_{12})$
coincides with properly  normalized matrix elements  of the operators 
${\rm e}^{\pm{\rm i}\beta\phi(0)}\,\sigma^{\mp}$:
\begin{equation}
\label{sa}
\frac{\langle\, \Omega_2\,|\,{\rm e}^{\pm{\rm i}\beta\phi(0)}\, \sigma^{\mp}\, |\,\Omega_1\,\rangle}
{\sqrt{\langle \Omega_2|{\rm e}^{-{\rm i}\beta\phi(0)}\sigma^+  |\Omega_2\rangle
\langle \Omega_1|{\rm e}^{+{\rm i}\beta\phi(0)}\sigma^-  |\Omega_1\rangle}}=g_\xi(\alpha_{12}) \,.
\end{equation}

An immediate check is to study the behavior at large $T_K^{(2)}/T_K^{(1)}$, where  we find
% easily
%$
%g_\xi(\alpha)\sim \exp\left(|\alpha|\, \xi/ 4(\xi+1)\right) \ {\rm for} \ \alpha\to\pm\infty$
%%
%so that
%
\begin{eqnarray}
\label{asym}
\langle \Omega_2|\Omega_1\rangle \simeq
\begin{cases} 
C_\xi\,  
\Big({T_K^{(2)}\over T_K^{(1)}}\Big)^{-{\xi\over 4(\xi+1)}}
&(\xi<\infty)\\
C_{\infty}\,  
\Big( \log\frac{T_K^{(2)}}{T_K^{(1)}}\Big)^{\frac{3}{4}}\ \Big({T_K^{(2)}\over T_K^{(1)}}\Big)^{-{1\over 4}}
   &(\xi=\infty)
\end{cases}
%  \   \mbox{ for }\    T_K^{(1)}\gg T_K^{(2)} \,.
\nonumber\\
\label{overlapasymptotics}
\end{eqnarray}
for $T_K^{(2)}\gg T_K^{(1)}$.
This is in agreement with the dimension of the boundary condition changing operator for 
the anisotropic Kondo problem, $d_K={1\over 4}{\xi\over \xi+1}$.
For the isotropic Kondo case ($\xi = \infty$) we recover the dimension, 
$d_K = \frac14$, of the $j={1\over 2}$ $SU(2)$ primary. 
Notice that, since in  eq.\,\eqref{mainres} we assume 
the conventional  normalization condition $\langle\, \Omega\,|\,\Omega\,\rangle=1$,
the constants  in  asymptotic formulae \eqref{asym}  are universal amplitudes.  Their expression can be found in the supplementary material. 
Of course, one can also verify that~\eqref{mainres} is consistent 
with the perturbative results~\eqref{ratioresi} and~\eqref{ratiores}.

  {\it Numerical results. }We now turn to a detailed numerical exploration of our result. There 
are various lattice models where the overlap~\eqref{mainres} can be measured. 
We have focused on the XXZ chain with a weak boundary coupling
\begin{equation}
H=\sum_{i=0}^{N} t_i\left(S^x_{i}S^x_{i+1}+S^y_i S^y_{i+1}+\Delta S_i^z S_{i+1}^z\right),
\end{equation}
where $t_i=1$ for $i\neq 0$ and $t_0=J_\ell$.
Standard bosonization~\cite{AffleckXXZ}
shows that this Hamiltonian is equivalent, at low energy, to~(\ref{K1}) with the 
Kondo coupling $J\propto J_\ell$ and ${\beta^2\over 8\pi}=1-\frac{1}{\pi}\, 
\hbox{arccos }\Delta$ \footnote{Note that the interaction term $J_\ell \Delta S_0^z S_{1}^z$ on 
the weak link is marginal, but can nevertheless be ignored in~(\ref{K1}) 
since it appears with amplitude $J_\ell \to 0$ in the scaling limit.}.
From a numerical point of view, the easiest case to check is of course the 
Toulouse point where $\Delta=0$, for which the overlap~\eqref{eqUniversalFunction} can be expressed as 
a determinant of a matrix whose size scales linearly with the number of sites (see  e.g.~\cite{EchoCFT}).
 Results are presented in Fig.\,\ref{fig:Numerics}(a).
 While the agreement with the theoretical value is clearly good  -- note that there is no free 
parameter in~\eqref{mainres} -- several aspects are important to notice. 
First, the overlap varies very slowly with the ratio of Kondo temperatures. 
This requires exploring ratios $T_K^{(2)}/T_K^{(1)}$ of the order of $10^2$ or more. 
Since  the analytical result is only true in the scaling limit where $J_\ell \ll 1$, this forces us to 
explore extremely small values of the bare coupling. 
For these values, the Kondo screening length $1/T_K \propto (J_\ell)^{-2}$ is in turn very large. 
To avoid finite size effects -- which seem quite important for the determination of the overlaps -- we finally have to study 
larger systems than one would have expected -- of the order of $10^4$ sites, 
forbidding us in particular from testing the region where the overlap becomes very small. 
%Luckily, in this region, the behavior is entirely controlled by the orthogonality exponent, whose value is
%simply given by (\ref{overlapasymptotics}).

 The interacting case requires use of the DMRG technique~\cite{White}. 
We use here a two-site version in the matrix product state 
(MPS) language~\cite{Schollwoeck11}. In this case, we have been limited 
to chains of about 800 sites, for which finite size effects in the scaling limit remain unfortunately important. 
In order to obtain usable results, we have had to perform a {\sl double extrapolation}. 
For finite, small $J_\ell$ we have first extrapolated results for different  sizes to $N=\infty$. 
These results are represented in Fig.\,\ref{fig:Numerics}(b)
for $\xi = 1/3$. We have then performed a  second extrapolation  for different values of $J_\ell$ to $J_\ell=0$, represented by the black symbols in the figure. The result of these extrapolations is found to be consistent with the analytical result~\eqref{mainres}. 
Note that in principle, one would also need to extrapolate the bond dimension $\chi$ of the variational MPS used in DMRG to infinity, but we find that keeping $\chi \sim 100-300$ was enough  
for the finite $\chi$ effects to be negligible compared with 
the more important finite $N$ and finite $J_\ell$ effects. 

 {\it Discussion. } It is clear a posteriori -- in view of its extremely slow variation  with the ratio 
of Kondo temperatures -- that the overlap in the crossover would be impossible to obtain perturbatively. 
It is also  difficult to measure it numerically. 
The slow variation quantifies the weak dependency of
 the Kondo ground state on the impurity coupling. 
It would be interesting to obtain a more qualitative  understanding of this effect in terms of 
the screening cloud. Technically, the exact formula for the ground states 
overlap is the building stone for the calculation of general overlaps between quantum 
impurity systems with different boundary conditions. 
Exact calculations of Loschmidt echoes and work distributions in quantum quenches then follow using more traditional techniques~\cite{BCCFF1}, which will be discussed elsewhere. 
 
Despite their importance in the context of quantum information, the thermodynamic limit of similar ground state overlaps (fidelities) remain extremely difficult to access exactly -- even for non-interacting systems -- and are often non-perturbative in the relevant expansion parameters. Our result opens the door to the calculation of such overlaps in integrable systems.

\smallskip

\paragraph{Acknowledgments.}
We thank I.~Affleck, N.~Andrei, J.~Dubail, F.H.L.~Essler, A.M.~Tsvelick and A.B.~Zamolodchikov for useful discussions.  
HS gratefully acknowledges the hospitality of the Rutgers Physics Department where this work was started.  
The work of HS and JJ was supported by  the French Agence Nationale pour la Recherche (ANR Projet 2010 Blanc SIMI 4: DIME); the work of HS was also supported by the DOE  under grant number DE-FG03-01ER45908.  The work of RV was supported by the Quantum Materials program of LBNL, and  the work of SL was supported by the NSF  under grant number NSF-PHY-1404056.

\vfill
\eject

\clearpage
\newpage

\onecolumngrid

\centerline{\bf Supplementary material}

\subsection{A. Quantum Jost operators and  overlaps}

Here we present some details of the derivation of \eqref{mainres}.
Integrability of the Kondo problem has most often been used to find exact
wave functions with the Bethe ansatz. 
While this  should, in principle, give access to quantities such 
as the ground states overlap, 
this would require many  technical steps. 
We shall proceed here  differently, 
by exploiting integrability directly at the field theory level in a spirit of the seminal work \cite{Zam}.

We start from the Hamiltonian in the form (\ref{K1}) with an
infinitesimally small {positive}
magnetic field added
\begin{equation}
\label{K1supmat}
{\boldsymbol H} =H_0-h\,\sigma^z
+J \left({\rm e}^{{\rm i}\beta\phi(0)}\sigma^-+
{\rm e}^{-{\rm i}\beta\phi(0)}\sigma^+\right)\ ;\ \ \ 
\ \ \ \ \ H_0=\int_{-\infty}^{+\infty} {\rm d}x\, (\partial_x\phi)^2\ ,\ \ \ 
h\to 0^+\ .
\end{equation}
%
%where  the dimension of the perturbation is 
%${\beta^2\over 8\pi}= {\xi\over \xi+1}$.
Recall that 
%the orthogonality
%exponent $d_K=\frac{1}{4}\, \frac{\beta^2}{8\pi}$ and
the Kondo temperature varies (for constant $\xi\equiv\frac{\beta^2}{8\pi-\beta^2}$) as 
$T_K\propto J^{\xi+1}$.
Consider the interaction picture, treating the term  $\propto J$  as a perturbation.
Let us  perform a Wick rotation, $t\mapsto y={\rm i}\, t$, so that
for the Bose field in the interaction picture      $\phi(x,y)=\phi(x+{\rm i}y)$.
Introduce the $y$-ordered exponent
\begin{equation}
\label{Smatrix}
{\boldsymbol S}(y_2,y_1)={\cal T}_y\exp\left(- \int^{y_2}_{y_1}{\rm d}y\  J\,
\left({\rm e}^{{\rm }{\rm i}\beta\phi(0,y)}\,{\rm e}^{2h y}\sigma^-+
{\rm e}^{-{\rm i}\beta\phi(0,y)}\,{\rm e}^{-2h y}\ \sigma^+\right)\,
\right)
\end{equation}
and 
\begin{eqnarray}
\label{Lmatrix}
{\boldsymbol T}^{(-)}_J={\rm e}^{\frac{{\rm i} }{2}\beta\phi(0,0)\sigma^z}\ {\boldsymbol S}(0,-L)\ ,\ \ \ \ \ 
{\boldsymbol T}^{(+)}_J={\boldsymbol S}(+L,0)\ {\rm e}^{-\frac{{\rm i} }{2}\beta\phi(0,0)\sigma^z}\ ,
\end{eqnarray}
which are  operators acting on the impurity spin, whose matrix elements are themselves operators 
acting on the free bosonic degrees of freedom. 
Notice that  the conjugation condition $\big({\rm e}^{{\rm i}\beta\phi(0,y)}\,\sigma^-\big)^\dagger=
{\rm e}^{-{\rm i}\beta\phi(0,-y)}\,\sigma^+$ 
leads to the relation 
${\boldsymbol T}^{(+)}_J=\big({\boldsymbol T}^{(-)}_J\big)^\dagger$.
Take now the  adiabatic limit 
  $L\to+\infty$ 
assuming that the interaction is
switching on (off) adiabatically.
In the absence of interaction and for $h>0$,
the vacuum state for \eqref{K1supmat} is given by  the product 
$|\uparrow\,\rangle\otimes |\,{\rm vac}\,\rangle$, where
$|\,{\rm vac}\,\rangle$ stands for  the ground state  of 
$H_0$ and $|\uparrow\,\rangle$ is a spin-up eigenvector of $\sigma^z$.
With the common  physical intuition
it is  expected that the following limits exist and define
the so-called  Jost operators:
\begin{eqnarray}
\label{oapasps}
\lim_{h\to 0^+}{\rm lim}^{({\rm adiab)}}_{L\to+\infty}\ {\boldsymbol T}^{(-)}_J\,|\uparrow\,\rangle&=&
T_+(\alpha)\, |\,\uparrow\,\rangle+T_-(\alpha)\, |\downarrow\,\rangle\\
\lim_{h\to 0^+}{\rm lim}^{\rm(adiab)}_{L\to+\infty}\ 
\langle\, \uparrow\,|\,{\boldsymbol T}^{(+)}_J&=&
\langle\, \uparrow|\, T^\dagger_+(\alpha)+ \langle\, \downarrow|\, T^\dagger_-(\alpha)\ ,\nonumber
\end{eqnarray}
where we use the parameterization $J\propto {\rm e}^{-\alpha/(\xi+1)}$.

Let us introduce the complex coordinate $z=x+{\rm i}\, y $ and the polar angle $\psi=\arg({\rm i}\, z)$.
It is well known  that the generator of infinitesimal Euclidean rotations
coincides with $(-{\rm i}\, K)$, where  $K$ is
the Lorentz boost generator.
The Jost operators defined by  eq.\,\eqref{oapasps} are related
through  the Euclidean rotation   of angle $\pi$:
\begin{equation}
\label{conj}
T^{\dagger}_a(\alpha)={\rm e}^{-{\rm i} \pi K}\ T_{-a}(\alpha)\ {\rm e}^{+{\rm i} \pi K}\ .
\end{equation}
It is crucial for our analysis that
the  angular  evolution of $T_\pm$
is remarkably  simple:
\begin{equation}
\label{oapsaspas}
{\rm e}^{-{\rm i} \psi K}\ T_\pm(\alpha)\ {\rm e}^{+{\rm i} \psi K}=T_\pm(\alpha+{\rm i}\psi )\ .
\end{equation}
This   follows from  the fact that    exponential fields
in \eqref{Smatrix}
are chiral (holomorphic)  $\big({\rm e}^{\pm {\rm i}\beta\phi(x,y)}\equiv{\rm e}^{{\pm\rm i}\beta\phi}(z)\big)$
with the    Lorentz
spin $1-\frac{1}{\xi+1}$, while $J\propto{\rm e}^{-\alpha/(\xi+1)}$.
Also in the derivation of \eqref{oapsaspas},
we have to    accept that
integrals containing
combinations of  the {\it holomorphic} fields are not  affected by
rotation of the integration contour in the limit  $L\to\infty$.

The exact vacuum overlap $\langle\,\Omega_2\,|\,\Omega_1\,\rangle$ can be written as
\begin{equation}
\label{brak}
\langle\,\Omega_2\,|\,\Omega_1\,\rangle={}_+\langle\,\alpha_2\,|\,\alpha_1\,\rangle_+\, +\,{}_-\langle\,\alpha_2\,|\,\alpha_1\,\rangle_-\ ,
\end{equation}
where we use  bra and ket states
\begin{equation}
{}_{\pm}\langle\,\alpha\,|=\langle\,{\rm vac}\,|\,T^\dagger_\pm(\alpha)\ ,\ \ \ \ \ \ \ |\,\alpha\,\rangle_{\pm}=T_\pm(\alpha)\, |\,{\rm vac}\,\rangle\ .
\end{equation}
With  eqs.\,\eqref{conj} and \eqref{oapsaspas}, the formula\,\eqref{brak} takes  the form
\begin{equation}
\label{over2}
\langle\,\Omega_2\,|\,\Omega_1\,\rangle=\sum_{a=\pm}F_{-a\,a}(\alpha_1-\alpha_2-{\rm i}\pi)\ .
\end{equation}
Here we introduce
\begin{equation}
\label{functions}
F_{\pm \mp}(\alpha-\alpha')=\langle\,{\rm vac }\,|\, T_{\pm}(\alpha')\,T_{\mp}(\alpha)\,|\, {\rm vac}\,\rangle
\end{equation}
and use the fact that  these functions depend on the difference $\alpha-\alpha'$ only
(this  follows immediately from
the Lorentz invariance and eq.\,\eqref{oapsaspas}).

The model under consideration   does not possess $P$- and $T$-invariance. However the $C$-parity, 
\begin{equation}
C\ :\ \ \ \phi\mapsto -\phi\ ,\ \ \ \ \sigma^{\pm}\mapsto \sigma^{\mp}\ ,\ \ \ \sigma^z\mapsto -\sigma^z\ ,
\end{equation}
is an obvious symmetry of the Hamiltonian \eqref{K1supmat} 
with $h=0$. Despite the infinitesimally small  field $h$ that was  involved  in  our construction,
we should keep in mind  that the $C$-parity is not spontaneously broken for any $J\not =0$.
Among others  this implies an absence of spontaneous magnetization (the Kondo spin is totally screened), and
more 
generally
\begin{equation}
\langle\,\Omega_2\,|\,\sigma^z\,|\,\Omega_1\,\rangle|_{h=0}=
{}_+\langle\,\alpha_2\,|\,\alpha_1\,\rangle_+\, -\,{}_-\langle\,\alpha_2\,|\,\alpha_1\,\rangle_-
=0\ .
\end{equation}
This yields a relation between  the free vacuum expectation values   \eqref{functions}:
\begin{equation}
\label{pasaos}
F_{ +- }(\alpha)=F_{ -+ }(\alpha)\equiv f(\alpha)\ .
\end{equation}
The  vacuum overlap \eqref{over2}
can be  written now as
\begin{equation}
\label{pasap}
\langle\,\Omega_2\,|\,\Omega_1\,\rangle=2\, f(\alpha_{12}-{\rm i}\pi)\ \ \ \ \ \ \ \ \ \ \  \ 
(\alpha_{12}\equiv \alpha_1-\alpha_2)\ ,
\end{equation}
and since $\langle\,\Omega_2\,|\,\Omega_1\,\rangle=\langle\,\Omega_1\,|\,\Omega_2\,\rangle$
\begin{equation}
\label{pasaod}
f(\alpha-{\rm i}\pi)=f(-\alpha-{\rm i}\pi)\ .
\end{equation}
Notice also  that we always assume  that   $\langle\,\Omega\,|\,\Omega\,\rangle=1$ and  therefore
\begin{equation}
\label{normal}
f(-{\rm i} \pi)=\frac{1}{2}\ .
\end{equation}

It is expected that the Jost operators  $T_\pm$ satisfy  remarkable commutation relations,
which follow from the 
canonical commutator  $[\phi(0,y_2),\,\phi(0,y_1)]=\frac{{\rm i}}{4}\ {\rm sign}(y_2-y_1)$
for the fields $\phi(0,y)$, appearing    in the definitions \eqref{Smatrix} and \eqref{Lmatrix}.
As in Ref.\,\cite{BLZ} one can show that,
acting on a pair of spins by two matrices 
${\boldsymbol T} ^{(-)}_J$ and ${\boldsymbol T} ^{(-)}_{J'}$, the Yang Baxter equation is satisfied in the form:
\begin{equation}
{\check {\boldsymbol R}}\ \big(\,{\boldsymbol T}^{(-)}_J\otimes 1\,\big)\,
\big(\,1\otimes {\boldsymbol T}^{(-)}_{J'}\,\big)=
\big(\,1\otimes {\boldsymbol T}^{(-)}_{J'}\,\big)\, 
\big(\,{\boldsymbol T}^{(-)}_{J}\otimes 1\,\big)\ {\check {\boldsymbol R}}\, . \label{YB}
\end{equation}
Nontrivial matrix elements of the  $4\times 4$ matrix ${\check {\boldsymbol R}}$ read explicitly
 \begin{eqnarray}
 {\check R}_{++}^{++}={\check R}_{--}^{--}=1\,,\ \ \ \ \ \ \ \
%\nonumber\\
  {\check R}_{+-}^{+-}={\check R}_{-+}^{-+}=- {\sinh {\alpha\over \xi+1}\over \sinh {i\pi+\alpha\over \xi+1}}\, ,\ \ \ \ \ \ \ \ 
%\nonumber\\
    {\check R}_{+-}^{-+}={\check R}_{-+}^{+-}= {\sinh {i\pi\over \xi+1}\over\sinh{i\pi+\alpha\over\xi+1}}\,,
    \end{eqnarray}
where   $\alpha=-(\xi+1) \log( J/J')$.
At this point, we still do not know how to perform the adiabatic limit on a rigorous base.
However, arguments similar to those from
Ref.\,\cite{Sergei1} suggest   that the Jost operators satisfy the  commutation relations
\begin{equation}
\label{ZF}
T_a(\alpha_1)\,T_b(\alpha_2)=\sum_{c,d=\pm} R_{ab}^{cd}(\alpha_1-\alpha_2) \, T_d(\alpha_2)\,T_c(\alpha_1)\ ,
\end{equation}
where $R_{ab}^{cd}(\alpha)=R(\alpha)\, {\check R}_{ab}^{cd}(\alpha)$. 
%This equation, while compatible with (\ref{YB}), is simpler since it involves the $R$-matrix on one side only. 
The self-consistency of  these relations for $a=b$  requires that
the normalization factor $R(\alpha)$  obeys the condition $R(\alpha)R(-\alpha)=1$. It is found to be 
\begin{equation}
R(\alpha)=\exp\left({\rm i}\,\int_{0}^\infty 
{{\rm d} t\over t}\ 
\frac{\sin(2\alpha t/\pi)}{\cosh t}\ \frac{\sinh t\xi}{\sinh t (\xi+1)}\,
\right)\, .
\end{equation}
With this choice the $R$-matrix satisfies both
``unitarity'' and  ``crossing symmetry'' relations
\begin{equation}
\label{crsym}
\sum_{c, d=\pm} R_{ab}^{cd}(\alpha)R_{dc}^{ef}(-\alpha)=\delta^{f}_{a}\delta_b^e\ ,\ \ \ \ \ \ \ \ 
R_{ab}^{cd}(\alpha)=R_{b{\bar c}}^{d{\bar a}}(-{\rm i}\pi-\alpha)\, ,
\end{equation}
where ${\bar c}\equiv -c,\ {\bar a}\equiv-a$.

The formal algebraic  relations \eqref{ZF} imply that
the function $f(\alpha)$, defined by formulae \eqref{functions} and \eqref{pasaos}, obeys the condition
\begin{equation}
\label{iosasaisa}
\frac{f(-\alpha)}{f(\alpha)}=
\frac{\sinh\frac{{\rm i}\pi-\alpha}{2(\xi+1)}}{\sinh\frac{{\rm i}\pi+\alpha}{2(\xi+1)}}\ 
R(\alpha)\ .
\end{equation}
Let us assume now that  for real $\alpha_1$, the free vacuum expectation values
$F_{ \pm\mp }(\alpha_2,\alpha_1)$\  \eqref{functions}
being considered as  functions of the complex variable $\alpha_2$, are analytic
in the strip $0\leq \Im  m(\alpha_2)\leq \pi$, or equivalently, $f(\alpha)$ is an analytic function  for
$-\pi\leq \Im  m(\alpha)\leq 0$.
Because of eq.\,\eqref{pasap},
we should also expect that as   $\alpha\to\pm\infty$, 
$f(\alpha-{\rm i}\pi)\sim {\rm e}^{-d_K|\alpha|}$, where $d_K$
is some
(a priori unknown) exponent.
With these two analytical assumptions,   conditions  \eqref{pasaod},\,\eqref{normal} and \eqref{iosasaisa}
define   $f(\alpha)$    unambiguously.
The final  analytical expression   for the overlap   is quoted in our main result \eqref{mainres}.

Notice that similar  considerations   can be applied to  vacuum matrix elements  
of the operators ${\rm e}^{\pm{\rm i}\beta\phi(0)}\,\sigma^{\mp}$. This yields the following prediction
\begin{equation}
\label{saisa}
\langle\, \Omega_2\,|\,{\rm e}^{\pm{\rm i}\beta\phi(0)}\, \sigma^{\mp}\,  |\,\Omega_1\,\rangle=
B_\xi\ \big(\,T^{(1)}_K\,T^{(2)}_K\,\big)^{\frac{\xi}{2\xi+2}}\ g_\xi(\alpha_{12})\, ,
\end{equation}
where $g_\xi(\alpha)$ is  defined  by \eqref{gxi} 
and $B_\xi$ stands for  some  $\alpha$-independent constant.
Eq.\,\eqref{saisa} can be rewritten in the form \eqref{sa}  presented in the main body of the text.
The value of the  constant $B_\xi$ depends on the
normalization condition imposed on the operators as well as  on the  precise definition   of physical energy
scale -- the Kondo temperature.
The conventional  normalization is defined through the condition imposed on the leading  term of  operator product expansion
of the field ${\boldsymbol V}(y)\equiv {\rm e}^{+{\rm i}\beta\phi(0,y)}\,\sigma^-+
{\rm e}^{-{\rm i}\beta\phi(0,y)}\,\sigma^+$:
\begin{equation}
{\boldsymbol V}(y){\boldsymbol V}(0)\to {\boldsymbol I}\ y^{-\frac{2\xi}{\xi+1}}\ \ \ \ \ {\rm as}\ \ y\to 0^{+} \ ,
\end{equation}
where ${\boldsymbol I}$ stands for the identity operator. As for the Kondo temperature, it can be  introduced
through the low-temperature behavior of the heat capacity  of the impurity:
\begin{equation}
\lim_{T\to 0} C(T)/T=\frac{\pi}{6 T_K}\ .
\end{equation}
With these conventions the constant $B_\xi$ can be extracted from the results of Ref.\,\cite{BLZ}
\begin{equation}
B_\xi=
-\frac{1}{2}\ \tan\Big(\frac{\pi\xi}{2}\Big)\  \Gamma\Big(\frac{1}{\xi+1}\Big)\ 
\left(\frac{\Gamma(\frac{\xi}{2})}
{\sqrt{\pi}\Gamma(\frac{\xi}{2}+\frac{1}{2})}\right)^{\frac{1}{\xi+1}}\ .
\end{equation}

We recall meanwhile that another way to exploit integrability in the Kondo model is to use a basis of the free boson 
Hilbert space made of  right moving massless kinks/antikinks \cite{Warner}, 
which are specific combinations of plane waves 
that scatter {\sl without particle production} on the Kondo impurity.
Such states (which  are, in some sense,  Bethe states in the infinite-size system limit)   can be denoted as
\begin{equation}
\label{oaspsopsaposa}
|\,\theta_1,\theta_2,\,\ldots\theta_n;\Omega\,\rangle_{a_1 a_2\cdots a_n}\ .
\end{equation}
Here  the subscript $a_i=\pm$ labels the type of $i^{\rm th}$  particle (kink or antikinks),
whereas
$\theta_i$ is  a rapidity  parameterizing  its energy and momentum  as $e_i=p_i=M\,{\rm e}^{-\theta_i}$ with $M$
being  an arbitrary   mass scale.
As $\xi\geq 1$,
the    states  \eqref{oaspsopsaposa}  form an overcomplete linear set. To use them as a linear independent basis
in the Hilbert space of the free chiral boson, 
one should chose  some ordering of the rapidities, say $\theta_1<\theta_2<\ldots<\theta_N$.
The price to pay in doing this is that there is now a  non trivial  kink-kink ``scattering''. 
It is most conveniently expressed in terms of the Zamolodchikov-Faddeev  commutation relations 
\begin{equation}
Z_a(\theta_1)\,Z_b(\theta_2)=\sum_{c,d=\pm }S_{ab}^{cd}(\theta_1-\theta_2)\,Z_d(\theta_2)Z_c(\theta_1)\ .
\end{equation}
The $S$-matrix here is very similar to the $R$-matrix. In fact, it is obtained  through the formal 
substitution:
\begin{equation}
S_{ab}^{cd}(\theta)=R_{ab}^{cd}(\alpha)\big|_{\theta\mapsto-\alpha,\,\xi\mapsto -\xi-1}\ .
\end{equation}
Note that the  finite Lorentz boost transformation produces  an overall  shift  of   all  rapidities, i.e.
\begin{equation}
{\rm e}^{-\theta K}\,Z_a(\theta')\, {\rm e}^{\theta K}=Z_a(\theta'+\theta)\ .
\end{equation}
This is  similar to eq.\,\eqref{oapsaspas}, however
the meaning of $\theta$ is a priori very different from that of
$\alpha$ -- the variable which parameterize the Kondo temperature. In fact, the $\alpha$ variable can be  adjusted  so that
$T_K=M\,{\rm e}^{-\alpha}$. (The  mass scale $M$ disappears at the end of the calculations,
since physical properties in the scaling regime depend only on ratios such as $T/T_K, h/T_K$, etc.)
The complete set of commutation relations for the Jost and Zamolodchikov-Faddeev operators
includes also \cite{Sergei2}
\begin{equation}
\label{KondoSm}
Z_a(\theta)\,T_b(\alpha)=ab~ {\rm i}~\tanh\left({\theta-\alpha\over 2}-
{{\rm i}\pi\over 4}\right)\, T_b(\alpha)\,Z_a(\theta)\ ,
\end{equation}
where on the right-hand side we recognize the anisotropic Kondo reflection
(or transmission, in the unfolded point of view) matrix \cite{Fendley}.

As well as the vacuum overlaps discussed above,   the  overlaps  
\begin{equation}
\label{overl1}
{}_{a'_1 \ldots a'_m}\langle\,\Omega';\,\theta'_1,\ldots\theta'_m
|\,\theta_1,\ldots\theta_N;\Omega\,\rangle_{a_1 \cdots a_n}\ ,
\end{equation}
can be obtained in a closed analytical form.
Notice that our construction of   the Jost operators $T_\pm$ essentially employs
the fields  on the
half infinite line $-\infty<y< 0$ only.
For this reason, the problem
of   calculation of     general    overlaps
is most conveniently treated by the angular
quantization\,\cite{Sergei2}.
Let us illustrate this first for the vacuum overlap.
%-- the QFT counterpart of the Baxter corner transfer approach.

In the  Hamiltonian picture associated with the space  $-\infty<\log|z|< +\infty$, the polar angle $\psi$ plays the role of
(Euclidean) time, whereas the angular Hamiltonian coincides with $(-{\rm i}\,K)$. 
Let ${\cal H}$ be the Hilbert    space  associated with the ``equal time'' slice $\psi=0$,
and the Jost operators $T_\pm$
are understood  now as operators acting in this space. Then the free vacuum expectation values \eqref{functions}
can be represented by the traces over  ${\cal H}$, 
%(see e.g.\,\cite{Brazhnikov}):
\begin{eqnarray}
\label{oiapssa}
F_{\pm\mp}(\alpha_1-\alpha_2)=\langle\,{\rm vac}\,|\,  T_{\pm }(\alpha_2) \,T_\mp(\alpha_1)\,
|\, {\rm vac}\,\rangle=\frac{1}{Z_{\cal H}}\ {\rm Tr}_{\cal H}\left(\, 
{\rm e}^{{2\pi \rm i} K}\, T_{\pm}(\alpha_2)\, T_\mp(\alpha_1)
\right)
%=
%\frac{1}{Z_{\cal H}}\ {\rm Tr}_{\cal H}\left(\, {\rm e}^{2\pi {\rm i} K}\, T_{-a}(\alpha_2+{\rm i}\pi)\, 
%T_a(\alpha_1)\right) 
\end{eqnarray}
with  $Z_{\cal H}={\rm Tr}_{\cal H}\big(\,{\rm e}^{2\pi{\rm i} K}\, \big)$.
We may now observe that  the important condition \eqref{pasaod}  is just a manifestation
of the   cyclic property of  the trace.
Using the unitarity and  crossing symmetry relations  \eqref{crsym}
it is easy to check    that  the sum
$\sum_{a=\pm}T_{-a}(\alpha+{\rm i}\pi)\, T_{a}(\alpha)$ commutes with  $T_{a'}(\alpha')$
for any  $\alpha,\ \alpha'$ and $a'=\pm$. Furthermore, as it follows from \eqref{KondoSm},
this sum also commute with any  Zamolodchikov-Faddeev operator $Z_{\pm}(\theta)$.
For this reason it is natural to expect that the Jost operators acting  in the space ${\cal H}$, satisfy the
condition
\begin{equation}
\label{det}
T_-(\alpha+{\rm i}\pi)\, T_+(\alpha)+T_+(\alpha+{\rm i}\pi)\, T_-(\alpha)=1\ .
\end{equation}
Here the constant in the r.h.s. is 
dictated by the vacuum normalization  $\langle\,\Omega\,|\,\Omega\,\rangle=1$.
As well as $T_a$, the Zamolodchikov-Faddeev operators  can be understood as  operators
acting in the Hilbert space of angular quantization~\cite{Sergei2}. 
In this case they   satisfy an important additional  requirement
which is somewhat similar to \eqref{det}:
\begin{equation}
\label{pole}
Z_a(\theta_2)\, Z_b(\theta_1)=-\frac{{\rm i}\ \delta_{a+b,0}}{\theta_2-\theta_1- {\rm i}\pi}\ +\ {\rm regular \  \ as}\  
 \  \
\theta_2\to\theta_1+{\rm i}\,\pi\ .
\end{equation}

The formula for the vacuum overlap (i.e., eq.\,\eqref{over2} with $F_{\pm\mp}$  are  given by \eqref{oiapssa}) 
can be   generalized for any overlap   \eqref{overl1}, namely
\begin{equation}
\label{overl}
{}_{a'_1 \ldots a'_m}\langle\,\Omega';\,\theta'_1,\ldots\theta'_m
|\,\theta_1,\ldots\theta_n;\Omega\,\rangle_{a_1 \cdots a_n}
=\sum_{a=\pm}F_{-a'_m,\ldots, -a'_1,a_1,\ldots, a_n; -a,a}
(\theta_m'+{\rm i}\pi,\ldots\theta'_1+{\rm i}\pi, \theta_1,\ldots\theta_n;
\alpha'+{\rm i}\pi,\alpha)\, ,
\end{equation} 
where
\begin{equation}
\label{GFun}
F_{c_1, c_2, \ldots, c_N; b,a}(\theta_1,\ldots\theta_N;\alpha',\alpha)=
\frac{1}{Z_{\cal H}}\ {\rm Tr}_{\cal H}\big({\rm e}^{2\pi{\rm i} K}\ T_{b}(\alpha')\,
Z_{c_1}(\theta_1)\ldots Z_{c_N}(\theta_N)\, T_{a}(\alpha)\,\big)\ .
\end{equation}
Notice that as it follows from the Lorentz invariance,
\begin{equation}
F_{c_1,  \ldots c_N; b,a}(\theta_1+\theta,\ldots\theta_N+\theta;\alpha'+\theta,\alpha+\theta)=
F_{c_1,  \ldots c_N; b,a}(\theta_1,\ldots\theta_N;\alpha',\alpha)\ ,
\end{equation}
whereas  the unbroken $C$-parity leads to  additional  relations
\begin{equation}
\label{C-parity}
F_{c_1,\ldots c_N; b,a}(\theta_1,\ldots\theta_N;\alpha',\alpha)=
F_{-c_1,\ldots -c_N; -b,-a}(\theta_1,\ldots\theta_N;\alpha',\alpha)\ .
\end{equation}
Also 
the cyclic 
property of  the  trace, the  commutation  relations for   $T_\pm,\ Z_\pm$ and  $K$, 
as well as formulae \eqref{det} and \eqref{pole}
can be easily translated to    further  conditions  imposed  on  the functions \eqref{GFun}.
The whole set of functional relations
can be then  supplemented
by  certain   analytical requirements  similar to that
mentioned  in derivation of the vacuum overlap. 
Although  rigorous derivation of all these conditions  is lacking, 
one  can except them as ``axioms'' similar to what has been done for matrix 
elements to local observables in integrable massive QFT -- the so-called form-factor program~\cite{Smirnov}.

A constructive way to  calculate the functions \eqref{GFun} 
was proposed in Ref.~\cite{Sergei2}. 
It is based on  the free field representation of the algebra of  
operators $T_\pm,\ Z_\pm$ and  $K$ acting in the angular quantization space ${\cal H}$, 
similar to the  bosonization procedure of  
conformal algebras and their vertex operators  in  2D conformal field theory.
This approach provides  an explicit integral representation for \eqref{GFun}, 
which are  the building blocks for the overlaps \eqref{overl}.

%The appearance of two different $R$ matrices in the problem is profoundly related to the fact that 
%the quantum symmetries of the continuum Kondo theory and its 
%lattice regularizations such as the six-vertex model with inhomogeneous spectral parameters \cite{ReshSal} are different.

\subsection{B. Toulouse limit}

This section and the next provide perturbative checks of the main result \eqref{mainres}.
We start with the   Toulouse limit. 
In this case, the problem is well known to be equivalent to two 
non-interacting critical Ising models with a boundary magnetic field $h\propto J$ \cite{LLS}. 
We thus consider  the  single  boundary Ising model.

Let $y={\rm i}\, t$ be the imaginary time. As before, we are  going to  use  complex coordinates
$z=x+{\rm i} y$ and ${\bar z}=x-{\rm i} y$, so that $\partial_z\equiv\frac{1}{2}
\big(\frac{\partial}{\partial x}-{\rm i}\frac{\partial}{\partial y}\big),\, 
\partial_{\bar z}\equiv\frac{1}{2}
\big(\frac{\partial}{\partial x}+{\rm i}\frac{\partial}{\partial y}\big)$.
The scaling regime of
the critical (in a bulk $x<0$) Ising model   is governed  by
the  following  Euclidean action 
for  Majorana-Weyl  fermions $\psi$ and ${\bar\psi}$ \cite{GZ}:
\begin{equation}
{\cal A} = \frac{1}{2\pi{\rm i}}\ \int_{-\infty}^\infty {\rm d}y 
 \int_{-\infty}^0 {\rm d}x\, \left(\,\psi{\partial}_{\bar z}\psi-
\bar{\psi}\partial_z\bar{\psi}\,\right)+\frac{1}{2}\ 
 \int_{-\infty}^\infty {\rm d}y\,
 \left(\, \frac{1}{2\pi{\rm i} }\, (\psi\bar{\psi})|_{x=0}+a\dot{a}
\,\right)+ h\int_{-\infty}^\infty {\rm d}y\,
\sigma_B(y)\  ,
%{\rm i}\ a\, (\psi+\bar{\psi})|_{x=0}\ .
\label{bdryising} 
\end{equation}
where $a=a(y)$ ($\dot{a}\equiv{{\rm d}a\over {\rm d}y}$)
is a boundary fermionic degree
of freedom with (as it is  dictated by the  action)
two point function $\langle a(y) a(y') \rangle\to \frac{1}{2}\ {\rm sign}(y-y')$ as
$y\to y'$.
We have also used here the  ``boundary spin operator''
\begin{equation}
\label{sigmab}
\sigma_B={\rm i}\ a\, (\psi+\bar{\psi})|_{x=0}\ .
\end{equation}
%
%Note that the last term of (\ref{bdryising}) was multiplied by  the imaginary unit,
%as compared with \cite{GZ}, since we need the action to be {\sl real}. 
Extremizing the action gives two boundary equations:
\begin{eqnarray}
\label{ada}
(\psi-\bar{\psi})|_{x=0} = 4\pi h\, a\ ,\ \ \ \ \ 
\dot{a}= -{\rm i}\  h\, \left(\psi +\bar{\psi}\right)|_{x=0} \  .
\end{eqnarray}
%
%where we use the shortcut notations $\psi_B=\psi(x,y)|_{x=0},\, {\bar \psi}_B={\bar \psi}(x,y)|_{x=0}$.
Eliminating $a$ between these  relations leads to
\begin{equation}
{{\rm d}\over {\rm  d}y}\left(\psi-\bar{\psi}\right)|_{x=0}=- {\rm i}\,\lambda\, \left(\psi+\bar{\psi}\right)|_{x=0} \ ,
\label{elimadot}
\end{equation}
where we introduce the parameter $\lambda$ of the dimension of energy
\begin{equation}
\lambda= 4\pi h^2\ .
\end{equation}
%
%In the context
%of the Kondo problem, $\lambda$  can be understood as a Kondo temperature.

It is useful to expand the fermions in the bulk  in Fourier integrals:
\begin{eqnarray}
\psi(x,y) = 
\int_{-\infty}^\infty{\rm d}k\  a_k\ {\rm e}^{ {\rm i}z k}\ ,\ \ \ \ \ \ \ 
\bar{\psi}(x,y) &=& \int_{-\infty}^\infty{\rm d}k\  {\bar a}_k\ {\rm e}^{ -{\rm i}{\bar z} k}\ ,
\end{eqnarray}
where  the modes obey   anticommutation relations 
$ \{ a_k,a_{k'}\}=\{ {\bar a}_k,{\bar a}_{k'}\}=\delta_{k+k',0}$ and $\{ a_k,{\bar a}_{k'}\}=0$.
%  which follow immediately from the action \eqref{bdryising}.
%
%\begin{equation}
%\{A(\theta),A^\dagger(\theta')\}= \{{\bar A}(\theta),{\bar A}^\dagger(\theta')\}=\delta(\theta-\theta').
%\end{equation}
%
Then 
the above  boundary equations of motion  are
solved by  the conditions
\begin{eqnarray}
\bar{a}_k=R_B(k)\ a_k\ ,\ \ \ \ \  {\rm where  }\ \ \ \ \ \ \  \ 
R_B(k)= { k -{\rm i}\lambda\over  k+{\rm i} \lambda}\ ,
\end{eqnarray}
and
\begin{eqnarray}
\label{expn}
a(y)={\rm i}\ \sqrt{\frac{\lambda}{\pi}}\ \int_{-\infty}^\infty{\rm d} k\ 
\frac{a_k}{k+{\rm i}\lambda}\   \  {\rm e}^{-yk}\ .
%\\
%{\dot a}(y)&=&
%-{\rm i}\ \sqrt{\frac{\lambda}{\pi}}\int_{-\infty}^\infty{\rm d} k\ \frac{k\, a_k}{k+{\rm i}\lambda}\  \  {\rm e}^{-yk}\ ,
\nonumber
\end{eqnarray}
Using the contraction  ${\overbracket[.4pt]{a_{k}a}}_{k'}=
\frac{1}{2}\, (1+{\rm sign}\, k)\,  \delta_{k+k',0}$, it is easy to  calculate  
the two point function
of the boundary fermion in the presence of the field $h$:
\begin{eqnarray}
\label{GCOR}
\langle\, a(y)\,a(0)\,\rangle_h
={\rm sign}(y)\ \frac{\lambda}{\pi}\ \int_0^{\infty}\frac{{\rm d} k}{k^2+\lambda^2}\ {\rm e}^{-|y| k}\ .
%\\
%&=&
%\frac{1}{2}\, 
%{\rm sign}(Y)\  \cos(Y)+\frac{1}{\pi}\ \big(\,{\rm Ci}(|Y|)\sin (Y)-{\rm Si}(Y) \cos (Y)\,\big)\ \ \ \ 
%\ \ \  \ (Y= \lambda\, y)\ .\nonumber
\end{eqnarray}
%Here ${\rm Ci}(y)=-\int_{y}^\infty\frac{\rm d t}{t}\, \cos t$, ${\rm  Si}(y)=
% \int_0^y\frac{\rm d t}{t}\, \sin t$ 
%stand  for  common
% ``cosine'' and ``sine'' integrals, respectively.
Notice that this   scaling function
is known since McCoy's work \cite{McCoy}.

We can now find  the two point function of the  
boundary spin operator. Recall that $\sigma_B$ was defined by the formula \eqref{sigmab} and
therefore,
as follows from  the boundary equations \eqref{ada},
 \begin{equation}
\sigma_B(y)=\frac{1}{h}\  {\dot a} a(y)\ .
%= \sqrt{\frac{\lambda}{\pi}}\ 
%\int_{-\infty}^\infty{\rm d} k\,{\rm d} k'\ y\frac{ (k+ k')\,{\rm e}^{-y(k+k')} }{(k+{\rm i}\lambda)(k'+{\rm i}\lambda)}
%\ :a_ka_{k'}:\ .
\end{equation}
Using the  Wick theorem, one finds
\begin{equation}
\label{spinspin}
\big\langle\,\sigma_B(y)\,\sigma_B(0)\,\big\rangle^{\rm (conn)}_h=
%\equiv
%\big\langle\,\sigma_B(y)\sigma_B(0)\,\big\rangle_h-\langle\sigma_B\rangle^2_h=
\frac{1}{h^2}\ 
\big( G {\ddot G}-{\dot G}^2\big)\ ,
%\frac{4\pi}{\lambda}\ 
%\big(\, \langle a(y){\dot a}(0)\rangle_h\          \langle {\dot a}(y)a(0)\rangle_h-
%          \langle {\dot a}(y){\dot a}(0)\rangle_h\ \langle a(y) a(0)\rangle_h\, \big)\ .
 \end{equation}
where $G(y)$ stands for  the two point function 
%$\langle\, a(y)a(0)\,\rangle_h$
\eqref{GCOR} and  the  abbreviation ``conn''
means
the connected  correlation function, i.e.,
$
\langle\,\sigma_B(y)\,\sigma_B(0)\,\rangle^{\rm (conn)}_h
\equiv
\langle\,\sigma_B(y)\,\sigma_B(0)\,\rangle_h-\langle\sigma_B\rangle^2_h$.
%It is easy to check  that
% \begin{equation}
%\big\langle\,\sigma_B(y)\sigma_B(0)\,\big\rangle^{\rm (conn)}_h=
% \begin{cases}
% \frac{2}{| y|}+O\big(\log^2|y|\big)\ \ \ \ &{\rm as}\ \ \ y\to 0\\
%  \frac{4}{\pi\lambda^3  y^4}+O(y^{-6})\ \ \ \ &{\rm as}\ \ \ y\to \infty
%  \end{cases}
% \ .
% \end{equation}
The boundary magnetization 
$\langle\sigma_B\rangle_h=h^{-1} \langle {\dot a} a
\rangle_h$  is proved to be  a non-universal quantity depending on the
lattice spacing $\varepsilon$:
\begin{equation}
\label{paos}
\langle\sigma_B\rangle_h=4h\ \log\left(2h^2\varepsilon\right)\ .
\end{equation} 
%Notice that an immediate  consequence of eq.\,\eqref{paos}   is the well known 
%prediction  for  the  boundary  free  energy:
%\begin{equation}
%E(h)=-h^2\, \log\left(h^2\varepsilon/{\rm e}\right)\ :\ \ \ \ \ \ \langle\sigma_B\rangle_h=
%\partial_h E\\ .
%\end{equation} 

We now turn to  the partition function 
${\cal Z}(h,h+\delta )$ 
of an infinite system with boundary field $h$ everywhere except on 
a part of the boundary of length $\tau$ where the boundary field is taken to 
be $h+\delta $. 
Expanding the action in powers of $\delta $,
we obtain the first nontrivial   correction to the ratio of partition functions
\begin{equation}
{{\cal Z}(h,h+\delta )\over {\cal  Z}(h,h)}=
\Big(\,1+\delta^2\  {\cal J}(\tau,h)+O(\delta ^3)\,\Big)\, 
\exp\big(-\tau \delta \,\langle\sigma_B\rangle_h\,\big)\ ,
\end{equation}
where  ${\cal J}$ results from two integrations on the modified boundary
\begin{equation}
{\cal J}(\tau,h)=\int_0^\tau {\rm d}y_1\int_0^{y_1} {\rm d} y_2\  
\big\langle\, \sigma_B(y_1)\, \sigma_B(y_2)\,\big\rangle^{\rm (conn)}_h\ .
\label{calJintegral}
\end{equation}
In order to perform the integration here it is convenient to represent the
connected spin spin correlator \eqref{spinspin} as 
\begin{equation}
\big\langle\, \sigma_B(y_1)\,\sigma_B(y_2)\,\big\rangle^{\rm (conn)}_h=\frac{4\lambda}{\pi}\ 
\int_0^\infty\int_{0}^\infty {\rm d} m_1 {\rm d} m_2 \ 
\frac{m_1(m_1-m_2)}{(m_1^2+1)(m_2^2+1)}\ \ {\rm e}^{-\lambda |y_1-y_2|(m_1+m_2)}\ .
\end{equation}
The imaginary time integrals in (\ref{calJintegral}) affect only  
the term ${\rm e}^{-\lambda |y_1-y_2|(m_1+m_2)}$. 
They give a term linear in  $\tau$, 
and a term  $\propto {\rm e}^{-\lambda\tau(m_1+m_2)}$,
% (see after (\ref{imagtimeprop})),
plus the term of order one that we are after. A simple calculation yields
\begin{equation}
{\cal J}(\tau,h)=\chi_h\,\tau-\frac{1}{2\pi^2 h^2}\ \int_0^\infty\int_{0}^\infty {\rm d}m_1
{\rm d} m_2\  {(m_1-m_2)^2\over (m_1+m_2)^2 (m_1^2+1)(m_2^2+1)}+
O(\tau^{-2})\ \ \ \ {\rm as}\ \ \ \tau\to \infty\ .
\end{equation}
The double integral in the r.h.s.
equals one. Also the linear  slop  $\chi_h=-
\frac{1}{2} \frac{{\rm d}}{{\rm d} h}\langle\sigma_B\rangle_h.$
%$(2h)^{-1}\,
%\langle\sigma_B\rangle_h-2$. 
Finally, combining everything together,
we  find
\begin{equation}
{{\cal Z}(h,h+\delta )\over {\cal Z}(h,h)}\bigg|_{\tau\to+\infty}\simeq\left(1-
\frac{1}{2\pi^2 }\ \delta^2/h^2 +
O\big(\delta^4\big)
\right)\ {\rm e}^{ -\tau\Delta E}
  \ , \label{rescalc}
\end{equation}
where $\Delta E=\int_{h}^{h+\delta}{\rm d}h\ \langle\sigma_B\rangle_h$.
%\equiv E(h+\delta )-E(h)= \langle\sigma_B\rangle_h\, \delta -
% (\langle\sigma_B\rangle_h-4 h)/(2h)
%\chi_h\, \delta ^2+
%O(\delta ^3)$.
This concludes the derivation of (\ref{ratioresi}),
since  $h^2\propto {\rm e}^{\alpha}$, i.e., $ \delta/h \simeq \frac{1}{2}\, \alpha_{12}$.

\subsection{C. Semiclassical limit}

We next consider the overlap in the   semiclassical limit. It is convenient to use
the Hamiltonian (\ref{semiclass}),
\begin{equation}
{\boldsymbol H}={1\over 2}\int_{-\infty}^0{\rm d}x\,  \big(\, (\partial_t\Phi)^2+(\partial _x\Phi)^2\,\big)
+J \sigma^x+{\beta\over 4}\, \partial_t \Phi_B\,\sigma^z\, .\label{semiclasssupmat}
\end{equation}
Here the boundary value of the field $\Phi$
is denoted by  $\Phi_B$.
We want to do perturbation in powers of $\beta$ around the ``classical'' case $\beta=0$. In this limit the spin takes the value $\sigma^x=-1$, and the free boson propagate   independently on $J$. In particular,
the unperturbed   imaginary time  two point function of ${\dot \Phi}_B\equiv \frac{{\rm d}}{{\rm d}y}\Phi_B$
reads
\begin{equation}
\big\langle\,  {\dot  \Phi}_B(y_1)\,{\dot \Phi}_B(y_2)\, \rangle_0=\frac{1}{\pi}\ \frac{1}{|y_1-y_2|^2}=
\frac{1}{\pi}\ \int_{0}^\infty {\rm d}k\ k\, {\rm  e}^{-k |y_1-y_2|}\  .
\label{imagtimeprop}
\end{equation}
Perturbative expansion of the partition function 
${\cal Z}(J,J')$ for  a system with coupling $J$ everywhere along the imaginary time axis except for a part where it is $J'$ gives
an extensive contribution (corresponding to a boundary free energy), an exponential contribution
(corresponding to excited states propagating along the boundary),
and a term of order one, leading to 
\begin{equation}
{{\cal Z}(J,J')\over {\cal  Z}(J,J)}\bigg|_{\tau\to+\infty}\simeq \left(\,1-{\beta^2\over 16\pi}\ 
\int_0^\infty {\rm d}k \  \frac{(J'-J)^2\ k}{(k+J)^2(k+J')^2}+O(\beta^4)\,\right)\ {\rm e}^{-\tau\Delta E}\   .
\end{equation}
The calculation of the integral is elementary and, using the parameterization $J\propto {\rm e}^\alpha$,
one obtains
\begin{equation}
{{\cal Z}(J,J')\over {\cal  Z}(J,J)}\bigg|_{\tau\to+\infty}\simeq
\left(\,1+{\beta^2\over 8\pi}\left(1-{\alpha-\alpha'\over 2}\coth{\alpha-\alpha'\over 2}\right)+
O(\beta^4)\, \right)\  {\rm e}^{-\tau\Delta E}\ ,
\end{equation}
which coincides with the result (\ref{ratiores}) given in the main text.

\subsection{D. Universal amplitudes}

We give here the expression of the constants appearing as pre-factors in  equation (\ref{overlapasymptotics}): 
\begin{eqnarray}
&&C_\xi=(\xi+1)^{\frac{3}{4}} 
\exp\left(-\int_0^\infty {{\rm d}t\over 4t} {\sinh t \over \cosh^2 t}\ {\sinh t\xi\over \sinh t(\xi+1)}\right)\nonumber\\
&&C_\infty=2^{-\frac{1}{6}}\, ({\rm e}\pi)^{\frac{1}{4}}\, A_G^{-3}=0.72209\ldots\ ,
\end{eqnarray}
where
$A_G$ is Glaisher's constant.
\end{document}